\documentclass[10pt,twocolumn]{article}
\usepackage{graphicx}
\usepackage{fullpage}
\usepackage{amssymb,amsfonts,amsmath}
\usepackage{algorithmic}
\usepackage{algorithm}
\usepackage{colonequals}
\usepackage{xspace}
\usepackage{xcolor}
\usepackage{booktabs}
\usepackage{secdot}
\usepackage[none]{hyphenat} % remove hyphenations over new lines.
\usepackage[numbers,square,comma,sort&compress]{natbib}

\newcommand{\argmax}{\operatornamewithlimits{argmax}}
\newcommand\Model{\ensuremath\mathcal{M}}

\begin{document}

% get rid of widows and orphans.
%\widowpenalty=1000
%\clubpenalty=1000

\title{\textbf{Network Archaeology: Uncovering Ancient Networks from Present-day Interactions}}

\author{Saket Navlakha and Carl Kingsford\\Department of Computer Science and Center for
Bioinformatics and Computational Biology\\University of Maryland, College Park USA\\\texttt{\{saket,carlk\}@cs.umd.edu}}

\maketitle

%==========================================================================
%==========================================================================
\begin{abstract}

Often questions arise about old or extinct networks. What proteins interacted
in a long-extinct ancestor species of yeast?  Who were the central players in
the Last.fm social network 3 years ago?  Our ability to answer such questions
has been limited by the unavailability of past versions of networks.  To
overcome these limitations, we propose several algorithms for reconstructing a
network's history of growth given only the network as it exists today and a
generative model by which the network is believed to have evolved. Our
likelihood-based method finds a probable previous state of the network by
reversing the forward growth model. This approach retains node identities so
that the history of individual nodes can be tracked.  We apply these algorithms
to uncover older, non-extant biological and social networks believed to have
grown via several models, including duplication-mutation with complementarity,
forest fire, and preferential attachment. Through experiments on both synthetic
and real-world data, we find that our algorithms can estimate node arrival
times, identify anchor nodes from which new nodes copy links, and can reveal
significant features of networks that have long since disappeared.

\end{abstract}
%==========================================================================

%==========================================================================
%==========================================================================
\section{Introduction}

% -- background on network growth

Many biological, social, and technological networks are the product of an
evolutionary process that has guided their growth. Tracking how networks have
changed across time can help us answer questions about why currently observed
network structures exist and how they may change in the
future~\cite{Hopcroft2004}. Analyses of network growth dynamics have studied
how properties such as node centrality and community structure change over
time~\cite{Hopcroft2004, Golbeck2007, Palla2007, Tantipathananandh2009}, how
structural patterns have been gained and lost~\cite{KumarKDD2006}, and how
information propagates in a network~\cite{LeskovecSDM2007}.

% -- general question and importance of question

However, in many cases only a static snapshot of a network is available without
any node-by-node or edge-by-edge history of changes. Biology is an archetypical
domain where older networks have been lost, as ancestral species have gone
extinct or evolved into present-day organisms. For example, while we do have a
few protein-protein interaction (PPI) networks from extant organisms, these
networks do not form a linear progression and are instead derived from species
at the leaves of a phylogenetic tree. Such networks are separated by millions
of years of evolution and are insufficient to track changes at a fine level of
detail. For social networks, typically only a single current snapshot is available
due to privacy concerns or simply because the network was not closely tracked
since its inception. This lack of data makes
understanding how the network arose difficult.

% -- forward growing models: counter an important possible objection

Often, although we do not know a network's past, we do know a general principle
by which the network supposedly grew forward. Several network growth models
have been widely used to explain the emergent features of observed real-world
networks~\cite{Barabasi1999, Vazquez2003, Ispolatov2005, Leskovec2005,
KumarKDD2006, Leskovec2007, LeskovecKDD2008}.  These models provide an
iterative procedure for growing a network so that the randomly grown network
exhibits similar topological features (such as the degree distribution and
diameter) as a class of real networks. For example, \emph{preferential
attachment} (PA) has explained many properties of the growing World Wide
Web~\cite{Barabasi1999}. The \emph{duplication-mutation with complementarity}
(DMC) model was found by~\citet{Middendorf2005} to be the generative model that
best fit the \emph{D.~melanogaster} (fruit fly) protein interaction network.
The forest fire (FF) model was shown~\cite{Leskovec2005} to produce networks
with properties, such as power-law degree distribution, densification, and
shrinking diameter, that are similar to the properties of real-world social
networks.  Although these random graph models by themselves have been useful
for understanding global changes in the network, a
randomly grown network will generally not isomorphically match a target network. Hence, forward
growth of random networks can only explore properties generic to the model and
cannot track an individual, observed node's journey through time.

% our idea: added to help explain why models can be validated using our approach.

This problem can be avoided, however, if instead of growing a random network
forward according to an evolutionary model, we decompose the actual observed
network \emph{backwards} in time, as dictated by the model. The resulting
sequence of networks constitute a model-inferred history of the
present-day network.

% justification of our question

Reconstructing ancestral networks has many applications.  The inferred
histories can be used to estimate the age of nodes, and to track the emergence
of prevalent network clusters and motifs~\cite{Milo2002}.  In addition,
proposed growth models can be validated by selecting the corresponding history
that best matches known histories or other external information. Leskovec et
al.~\cite{LeskovecKDD2008} explore this idea by computing the likelihood of a
model based on how well the model explains each observed edge in a given
complete history of the network. This augments judging a model on its ability
to reproduce certain global network properties, which by itself can be
misleading. As an example, Middendorf et al.\@~\cite{Middendorf2005} found that
networks grown forward according to the small-world model~\cite{Watts1998}
reproduced the small-world property characteristic of the
\emph{D.~melanogaster} PPI network, but did not match the true PPI network in
other aspects. Leskovec et al.~\cite{Leskovec2005} made a similar observation
for social network models.  Ancestor reconstruction also can be used to
down-sample a network to create a realistic but smaller network that preserves
key topological properties and node labels.  This can be used for faster
execution of expensive graph algorithms or for visualization purposes.  In the
social network setting, if a network's owner decides to disclose only a single
network, successful network reconstruction would allow us to estimate when a
particular node entered the network and reproduce its activity since being a
member. This could have privacy implications that might warrant the need for
additional anonymization or randomization of the network.

%Previous studies have examined various properties of ancestral networks.

Some attempts have been made to find small ``seed graphs'' from which
particular models may have started.  \citet{Leskovec2007}, under the Kronecker
model~\cite{Kronecker}, and \citet{Hormozdiari07}, under a duplication-based
model, found seed graphs that are likely to produce graphs with specified properties.  These seed graphs can be thought of as the ancestral graphs at
very large timescales, but the techniques to infer them do not generalize to
shorter timescales nor do they incorporate node labels.
Previous studies of time-varying networks solve related network inference problems, but assume
different available data.  For example, the use of exponential random graph
models~\cite{Ahmed09,Guo07,Hanneke06} for  inferring dynamic networks requires
observed node attributes (e.g.\@ gene expression) at each time point. They are
also limited because they use models without a plausible biological mechanism
and require the set of nodes to be known at each time point. Other
techniques~\cite{Wiuf06,Mithani09} estimate the parameters of the growth model,
but do not reconstruct networks or do so by only modeling the loss and gain of edges
amongst a fixed set of nodes.
There has been some recent work on inferring ancestral biological networks
using gene trees~\cite{Dutkowski2007,Gibson2009}. These approaches
``play the tape'' of duplication instructions encoded in the gene tree
backwards. The gene tree provides a sequence-level view of evolutionary
history, which should correlate with the network history, but their relationship can also be complementary. Further, gene tree approaches can only capture node arrival
and loss (taken directly from the gene tree), and do not account for models of
edge evolution.
Network alignment between two extant species has also been used to find
conserved network structures, which putatively correspond to ancestral
subnetworks~\cite{Kelley2003,Flannick2006,Singh2007}. However, these methods do
not model the evolution of interactions, or do so using heuristic measures.

Finally, the study of ancestral biological \emph{sequences} has a long history,
supported by extensive work in phylogenetics~\cite{Felsenstein2003}. Sequence
reconstructions have been used to associate genes with their function,
understand how the environment has affected genomes, and to determine the amino
acid composition of ancestral life. Answering similar questions in the network
setting, however, requires significantly different methodologies.

% Our results

Here, we propose a likelihood-based framework for reconstructing predecessor
graphs at many timescales for the PA, DMC, and FF network growth models. Our
efficient greedy heuristic finds high likelihood ancestral graphs using only
topological information and preserves the identity of each node, allowing the
history of each node to be tracked. Using simulated data, we show that network
histories can be inferred for these models even in the presence of
some network noise.

When applied to a protein-protein interaction (PPI) network for
\emph{S.~cerevisiae} (baker's yeast), our reconstruction accurately estimates
the sequence-derived age of a protein when using the DMC model. Assuming either
the PA model~\cite{Barabasi1999} or the FF model~\cite{Leskovec2005} designed
for social networks results in a poorer estimate of protein age, which further
confirms DMC as a more reasonable model of the growth of PPI
networks~\cite{Middendorf2005}. The inferred, DMC-based history also identifies
functionally related proteins as the product of duplication events, estimates
the number of duplication events in which each protein is involved, and can
distinguish between core and peripheral protein complex members based on their
arrival time.

To compare the growth of biological networks with that of social networks, we
used our algorithms to generate an approximate order in which users joined the
Last.fm music social network. As expected, the DMC model does not extend well
to this domain, where PA performs best. The FF model also outperforms DMC in
identifying users who apparently mediated the network's growth by attracting
new members to join.

The ability of these algorithms to reconstruct significant features of a
network's history from topology alone further confirms the utility of models of
network evolution, suggests an alternative approach to validate growth models,
raises privacy concerns in social networks, and ultimately reveals that much of the history of a
network is encoded in a single snapshot.
%==========================================================================
%==========================================================================

%==========================================================================
%==========================================================================
\begin{raggedright}
\section{Network reconstruction algorithms}
\end{raggedright}
\label{sec:algs}

Suppose an observable, present-day network is the product of a growth process
that involved a series of operations specified by a model $\mathcal{M}$ (such
as preferential attachment).  The model $\mathcal{M}$ gives us a way to grow
the network forward. We see now how this process can be reversed to find a
precursor network.

We start with a snapshot of the network $G_t$ at time $t$, and would like to
infer what the network looked like at time $t - \Delta t$. One approach to find
the precursor network $G_{t-\Delta t}^{*}$ is to find the maximum \emph{a
posteriori} choice:
\begin{equation}\label{eqn:opt}
G_{t-\Delta t}^{*} \colonequals \argmax_{G_{t - \Delta t}}
\mathrm{Pr}(G_{t-\Delta t} \mid G_t,\mathcal{M},\Delta t)\,.
\end{equation}
In other words, we seek the most probable ancestral graph $G_{t - \Delta
t}^{*}$, given that the observed graph $G_t$ has been generated from it in time
$\Delta t$ under the assumed model $\mathcal{M}$. Our goal is to find an entire
most probable sequence of graphs $G_1,G_2,\cdots,G_{t-1}$ that led to the given
network $G_t$ under model $\mathcal{M}$.

Because the space of possible ancestral graphs grows exponentially with $\Delta
t$ for all reasonable models, Equation~\eqref{eqn:opt} poses a challenging
computational problem.  A heuristic simplification that makes inference somewhat
more feasible is to set $\Delta t = 1$ and greedily reverse only a single step
of the evolutionary model. While this will no longer find the maximum a
posteriori estimate for larger $\Delta t$, it is much more tractable.  Repeated
application of the single-step reversal process can derive older
networks. We make the first-order Markov model assumption (also made by the
growth models) that $G_t$ only depends on $G_{t-1}$.  In this case, applying
Bayes' theorem, we can rewrite Equation~\eqref{eqn:opt} as:
\begin{align}
G_{t-1}^{*} \colonequals& \argmax_{G_{t-1}} \frac{\Pr(G_t \mid
G_{t-1},\mathcal{M}) \Pr(G_{t-1} \mid \mathcal{M})}{\Pr(G_t \mid \mathcal{M})}
\\
=& \argmax_{G_{t-1}} \Pr(G_t \mid G_{t-1},\mathcal{M}) \Pr(G_{t-1} \mid
\mathcal{M}), \label{eqn:optsimp}
\end{align}
where the last equality follows because the denominator is constant over the
range of the $\argmax$. This formulation has the advantage that the model
$\mathcal{M}$ is being run forward as intended.  The formulation also has the
advantage that the prior $\Pr(G_t\mid \Model)$ in Equation~\eqref{eqn:optsimp}
can be used to guide the choice of $G_{t-1}$.  Computing $\Pr(G\mid\Model)$
exactly for various models is an interesting computational problem in its own
right~\cite{Bezakova06} with a number of applications beyond ancestral network
reconstruction.  For computational simplicity, here we assume a uniform prior
and therefore consider the term a constant.

The ancestral reconstruction algorithm chooses the
predecessor graph with the largest conditional probability $\Pr(G_t \mid
G_{t-1},\mathcal{M})$ by searching over all possible predecessors graphs,
$G_{t-1}$.  In all models we consider, a single new node enters the network in
each time step and connects to some existing nodes in the network.  In the DMC
and forest fire models, the new node performs a link-copying procedure from a
randomly chosen \emph{anchor node}.  Finding the most probable network
predecessor graph therefore corresponds to finding and removing the most
recently added node, identifying the node it duplicated from (if applicable to
the model), and adding or removing edges that were modified when the most
recently added node entered the network.  In the next sections, we explain how
to do these steps efficiently for the DMC, PA, and FF growth models.
%==========================================================================

%==========================================================================
\begin{raggedright}
\subsection{The duplication-mutation with complementarity (DMC) model}
\label{sec:dmc}
\end{raggedright}

The DMC model is based on the duplication-divergence principle in which gene
duplication produces a functionally equivalent protein, which is followed by
divergence when the pair specialize into subtasks.  \citet{Middendorf2005} and
\citet{Pereira-Leal2006} have provided support and an evolutionary basis for
the general duplication model, which has been widely studied as a route by
which organism complexity has
increased~\cite{Vazquez2003,Wagner2003,Ispolatov2005,Levy2008}.  Though some
questions remain about its exact role in evolution~\cite{Kim2008}, the DMC
model appears to have a computational and biological basis for reproducing many features of real
protein interaction networks.

\newcommand\qmod{\ensuremath q_{\textrm{mod}}}
\newcommand\qcon{\ensuremath q_{\textrm{con}}}

\begin{figure*}[t]
\centerline{\includegraphics[width=0.79\textwidth]{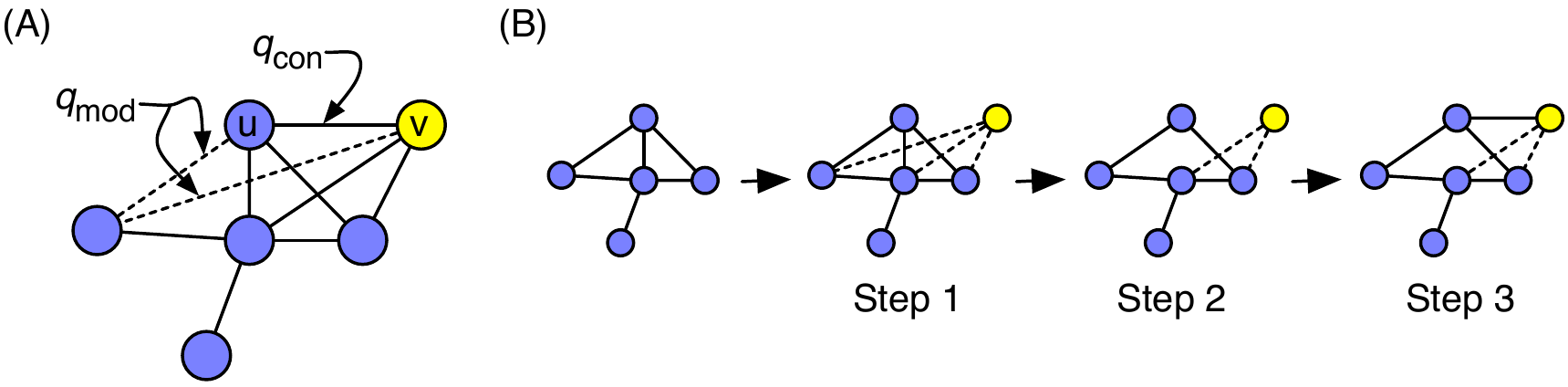}}
\caption{\label{fig:dmc-schematic}(A)~The probabilities governing the DMC model. (B)~An example iteration of the DMC model. Steps refer to those in Section~\ref{sec:dmc}.}
\end{figure*}

The forward DMC model begins with a simple, connected two-node graph. In each step,
growth proceeds as follows:
\begin{enumerate}
\item Choose a random anchor node $u$ and create its duplicate, $v$, by connecting $v$ to all of $u$'s neighbors.

\item For each neighbor $x$ of $v$, decide to modify the edge or its compliment
with probability $\qmod$.  If the edge is to be modified, delete either edge
$(v,x)$ or $(u,x)$ by the flip of a fair coin.

\item Add edge $(u,v)$ with probability $\qcon$.
\end{enumerate}
A schematic of the growth process is shown in Figure~\ref{fig:dmc-schematic}.

To reverse DMC, given the two model parameters $\qmod$ and $\qcon$, we attempt
to find the node that most recently entered the current network $G_t$, along
with the node in $G_{t-1}$ from which it duplicated (its anchor). Merging this
pair produces the most likely predecessor graph of
Equation~\eqref{eqn:optsimp}. Formally, $G_{t-1}$ is formed by merging:
\begin{align}
& \argmax_{(u,v)} \frac{\gamma_{uv}}{n} \prod_{N(u) \cap N(v)}\left(1-\qmod\right) \prod_{N(u)\bigtriangleup N(v)}\qmod, \label{eqn:dmc}
\end{align}
where $n$ is the number of nodes in $G_{t-1}$, $\gamma_{uv}$ equals $\qcon$ if
$u$ and $v$ are connected by an edge and $1-\qcon$ if not, $N(u)$ denotes
neighbors of node $u$, and the pairs $(u,v)$ range over all pairs of nodes in
$G_t$.  The expression inside the $\argmax$ of Equation~\eqref{eqn:dmc}
corresponds to $\Pr(G_t \mid G_{t-1},\mathcal{M})$, which is what we are trying
to maximize in Equation~\eqref{eqn:optsimp} by selecting $G_{t-1}$.  The $1/n$
factor gives the probability that node $u$ was chosen as the node to be
duplicated. The first product considers the common neighbors between the two
nodes. In the DMC model, a node and its duplicate ultimately share a neighbor
$x$ if $x$ was not modified in step~2 of the model.  The probability of such an
event is $1-\qmod$. The second product involves the nodes that are neighbors of
$u$ or $v$ but not both (symmetric difference of $N(u)$ and $N(v)$). Each such
neighbor exists with probability $\qmod$.

If $(u,v)$ is a pair that maximizes~\eqref{eqn:dmc}, the predecessor graph
$G_{t-1}$ is formed by removing either $u$ or $v$. Let $G^{vu}_{t-1}$
correspond to the graph where $v$ is removed. Due to symmetry, both
$G^{uv}_{t-1}$ and $G^{vu}_{t-1}$ yield the same likelihood in
Equation~\eqref{eqn:dmc}, and thus we are forced to arbitrarily decide which
node to remove.
Assume we randomly choose to remove
$v$; then $u$ gains edges to all nodes in $N(u) \cup N(v)$ that it does
not already have an edge to. This is because, according to the forward growth
model, $u$ originally had these edges prior to the duplication event of $v$ and
subsequent divergence.

Any pair of nodes in $G_t$ could correspond to the most recently duplicated
pair, including pairs with no common neighbors (which would happen if after
duplication all edges were modified in step~$2$ of the model). Thus, all
$\binom{n}{2}$ pairs of nodes must be considered in Equation~\eqref{eqn:dmc}.

%==========================================================================
\begin{raggedright}
\subsection{The forest fire (FF) model}
\label{sec:ff}
\end{raggedright}

The forest fire (FF) model was suggested by Leskovec et al.~\cite{Leskovec2005}
to grow networks that mimic properties of social networks. These
properties include power-law degree, eigenvalue, and eigenvector distributions,
community structure, a shrinking diameter, and network densification.

The forward FF model begins with a simple, connected two-node graph. In the
undirected case, in each step, growth proceeds according to the following
procedure with parameter $p$:
\begin{enumerate}

\item Node $v$ enters the network, randomly selects an anchor node $u$, and links to it.

\item Node $v$ randomly chooses $x$ neighbors of $u$ and links to them, where
$x$ is an integer chosen from a geometric distribution with mean $p/(1-p)$. These
vertices are flagged as active vertices.

\item Set $u$ to each active vertex and recursively apply step~$2$.  $u$
becomes non-active. Stop when no active vertices remain.

\end{enumerate}
To prevent cycling, a node cannot be visited more than once.  The
process can be thought of as a fire that starts at node $u$ and
probabilistically moves forward to some nodes in $N(u)$, then some nodes in
$N(N(u))$, etc.\@ until the spreading ceases. This version of the model only
contains one parameter: $p$, the burning probability. As in the DMC
model, the reversal process for the FF model attempts to find the node in the
current network $G_t$ that most recently entered the network, along with its
anchor.

Unfortunately, it appears to be difficult to write down an analytic expression
computing the likelihood of $G_{t-1}$. The main challenge is that for every $w \in
N(v)$ we need to find the likely paths through which the fire spread from
$u$ to $w$.
However, these paths are not independent, and therefore
cannot be considered separately. Analytic evaluation of
the global network properties produced by the model also appears to be
difficult~\cite{Leskovec2005}. Instead, we compute the likelihood of $G^{vu}_{t-1}$ via
simulation as follows:
\begin{quote}\raggedright
\textbf{Forest Fire Simulation Procedure.}
We assume $v$ does not exist in the network and simulate
the FF model starting from $u$. Each simulation produces a set of
visited nodes $S(v)$ corresponding to candidate neighbors of $v$. We
use the fraction of simulations in which $S(v)$ exactly equals $N(v)$
as the likelihood of $G^{vu}_{t-1}$.
\end{quote}

In the FF model, the likelihood of $G^{vu}_{t-1}$ does not necessarily equal
that of $G^{uv}_{t-1}$ because a forest fire starting at $u$ could visit
different nodes than a forest fire starting at $v$.  The advantage of
non-symmetry here is that there is no uncertainty regarding which node to
remove. Also, unlike the DMC model, all candidate node/anchor pairs must have
an edge between them (because of step~1 of the model).  After identifying
the node/anchor pair $v,u$ that yields the most likely $G_{t-1}$, we remove $v$
and all its edges from the graph. No edges need to be added to $u$ as per the
model.

Leskovec et al.~\cite{Leskovec2005} also propose a directed version of the FF model
where the fire can also spread to incoming edges with a lower probability.
Interestingly, reversing the directed FF model is much easier than the
undirected case because the node that most recently entered the network must
have exactly $0$ incoming edges. Choosing which of the nodes with a $0$
in-degree to remove first can be difficult because several nodes could have
been added to distant, independent locations in the graph in separate steps. A
node's anchor, however, can still be determined using our approach.

%==========================================================================
\begin{raggedright}
\subsection{The preferential attachment (PA) model}
\end{raggedright}

The preferential attachment (PA) model was proposed by Barab\'{a}si et
al.~\cite{Barabasi1999} in the context of growing networks to emulate the
growth of the Web. It follows the premise that new pages make popular pages
more popular over time by linking to them preferentially. We consider the
linear version of the PA model, which has been shown to correspond closely with
real network growth~\cite{LeskovecKDD2008}.

The PA model begins with a clique of $k+1$ nodes. In each step $t$, forward growth proceeds with parameter $k$ as follows:
\begin{enumerate}

\item Create a probability distribution histogram, where each node $u$ is assigned probability $d_u / (2m)$, where $d_u$ is the degree of $u$ and $m$ is the total number of edges in $G_{t-1}$.

\item Choose $k$ nodes according to the distribution.

\item Add node $v$, and link it to the $k$ nodes from step~$2$.
\end{enumerate}
Unlike the DMC and FF models, there is no notion of a node anchor in PA. A
new node simply enters the network in each step and preferentially attaches to
nodes with high degree. The most recently added node must be of minimum degree in $G_t$ because all nodes start with degree $k$ and can only gain edges over time. Let $\mathcal{C}$ be the set of nodes with minimum degree. To produce $G_{t-1}$, we choose a node to remove from among the nodes in $\mathcal{C}$ by computing:
\begin{align}\label{eqn:pa1}
& \argmax_{v \in \mathcal{C}} \prod_{u \in G_{t-1}} \begin{cases}
													d_u/m & \quad\text{if $u \in N(v)$}\\
													1-d_u/m & \quad\text{if $u \not\in N(v)$}
\end{cases}\,.
\end{align}
The two cases in the product correspond to whether edge $(v,u)$
exists.
The degree of $u$ in
$G_{t-1}$ can vary depending on which candidate node $v$ is being considered
for removal from $G_t$. Taking logs and simplifying turns \eqref{eqn:pa1} into:
\begin{align}
& \argmax_{v \in \mathcal{C}} \sum_{u \in G_{t-1}} \begin{cases}
													\log d_u - \log m & \text{if $u \in N(v)$}\\
													\log (m- d_u) - \log m & \text{if $u \not\in N(v)$}
\end{cases}\label{eqn:pa2}\\
=& \argmax_{v \in \mathcal{C}} \sum_{u \in N(v)} \log d_u + \sum_{u \not\in N(v)} \log(m-d_u)  \label{eqn:pa}
\end{align}
The $\log m$ terms in Equation~\eqref{eqn:pa2} can be ignored because they sum to $n\log m$ which is a constant over all candidate nodes. Equation~\eqref{eqn:pa} seeks to remove the node with
minimal degree that links to the ``hubbiest'' set of nodes. The likelihood is independent of
$k$.

%==========================================================================
%==========================================================================
\begin{raggedright}
\subsection{The reconstruction algorithms}
\label{sec:alg}
\end{raggedright}

\newcommand\Lpa{\ensuremath L_{\textrm{PA}}}
\newcommand\Ldmc{\ensuremath L_{\textrm{DMC}}}
\newcommand\Lff{\ensuremath L_{\textrm{FF}}}

The expression inside of the $\argmax$ of Equation~\eqref{eqn:dmc} for DMC defines a score for pairs of nodes. The corresponding score for PA
is given in Equation~\eqref{eqn:pa} and for FF in the simulation procedure. These scores
corresponds to the conditional probability $\Pr(G_t \mid G_{t-1},\mathcal{M})$
for each model. Let $\Ldmc(u,v)$, $\Lpa(u)$, and $\Lff(u,v)$ denote these
computed scores. To reverse each model, we iteratively search for the nodes
that maximize these scores. If there are ties, we randomly choose among them.
We continue this process until only a single node remains in the graph.  For
example, Algorithm~\ref{alg:greedy} gives the pseudocode for reversing a
network using the DMC model. The algorithm takes a static, present-day graph
$G=(V,E)$ and values for parameters $\qmod$ and $\qcon$.

\newcommand\Llist{\ensuremath P_{\textrm{list}}}
\newcommand\Lbest{\ensuremath L_{\textrm{best}}}

\begin{algorithm}
\algsetup{indent=2em}
\begin{algorithmic}[1]
\caption{\texttt{ReverseDMC(G = (V,E),$\qmod, \qcon$)}}
\label{alg:greedy}
\STATE \texttt{Arrival $\leftarrow \{\ \}$} \ \# Arrival time for each node
\STATE \texttt{Anchor $\leftarrow \{\ \}$} \# Anchor for each node
\WHILE{\texttt{|V| $\geq$ 2}}
\STATE \texttt{L$_{\texttt{best}}$ $\leftarrow$ -1;} \ \ \texttt{P$_{\texttt{list}}$ $\leftarrow$ [\ ]}
\FORALL{\texttt{pairs of nodes u,v $\in$ G}}
\STATE \texttt{L $\leftarrow$ L$_{\texttt{DMC}}$(u,v)}
\IF{\texttt{L = L$_{\texttt{best}}$}}
\STATE \texttt{insert (u,v) into P$_{\texttt{list}}$}
\ELSIF{\texttt{L > L$_{\texttt{best}}$}}
\STATE \texttt{P$_{\texttt{list}}$ $\leftarrow$ [(u,v)]}; \ \ \texttt{L$_{\texttt{best}}$ $\leftarrow$ L}
\ENDIF
\ENDFOR
\STATE \texttt{Choose a pair (u,v) from P$_{\texttt{list}}$ uniformly at random}
\STATE \texttt{Set Anchor[v] $\leftarrow$ u}
\STATE \texttt{Set Arrival[v] $\leftarrow$ |V|}
\STATE \texttt{Add edges (u,x) $\forall$x $\in$ N(v)-N(u) to E} \label{alg:removeedges}
\STATE \texttt{Delete v from G}
\ENDWHILE
\RETURN \texttt{(Arrival, Anchor)}
\end{algorithmic}
\end{algorithm}

Algorithm~\ref{alg:greedy} must be changed slightly for the FF and PA models.
For the FF model, the differences are: (1)~$\Lff(u,v)$ is used instead of
$\Ldmc(u,v)$; and (2)~the for-loop is over all pairs of nodes connected by an edge.
For the PA model: (1)~$\Lpa(u)$ is used; and (2)~the for loop is over all nodes
instead of all pairs of nodes. For both FF and PA no new edges are added to $v$
after node $u$ is deleted.

%==========================================================================
%==========================================================================
\begin{raggedright}
\section{Methods for validating reconstructed networks}
\label{sec:valid}
\end{raggedright}

\newcommand\truearr{\ensuremath A_{\textrm{true}}}
\newcommand\predarr{\ensuremath A_{\textrm{pred}}}
\begin{raggedright}
\subsection{Validating node arrival times}
\label{sec:nodearr}
\end{raggedright}
Our reconstruction framework gives an ordered list of node arrival times, with
the first removed node corresponding to the node that most recently entered.
Let $\truearr$ be the true arrival order of the nodes and let $\predarr$ be the
computationally predicted sequence. To understand how well our reconstructed
arrival times match the true node arrival times, we compute the difference
between $\truearr$ and $\predarr$ using the popular Kendall's $\tau$ and
Spearman's footrule measures~\cite{Bar-Ilan2006}:\\

\noindent \textbf{Kendall's tau}: $K_\tau = (n_c - n_d)/\binom{n}{2}$, where
$n_c$ is the number of concordant pairs in $\predarr$, i.e.\@ the number of
pairs in $\predarr$ that are in the correct relative order with respect to
$\truearr$; and $n_d$ is the number of discordant pairs. $K_\tau = 1$ if the
two lists are identical, and -1 if they are exactly opposite.\\

\noindent \textbf{Spearman's footrule:} $SF' = \sum_i |\truearr(i)
- \predarr(i)|$.  $A(i)$ is the node arrival time for node $i$.  This measure
  takes into account how far apart the arrival times are for each node in the
two lists. $SF'$ has a maximum value of $\lfloor n^2/2\rfloor$. We use a
normalized value of $SF = 1 - SF' / \lfloor(n^2/2)\rfloor$, so that $SF = 1$ if
the two lists are identical, and $0$ if they are opposite of each other.

In both cases, the higher the value the better.  The expected $K_\tau$ and $SF$
similarity between $\truearr$ and a random ordering of the nodes is $0.00$ and
$0.33$, respectively.

%==========================================================================
\begin{raggedright}
\subsection{Validating node anchors}
\label{sec:nodeanchors}
\end{raggedright}

\begin{figure}
\centering
\includegraphics[width=\columnwidth]{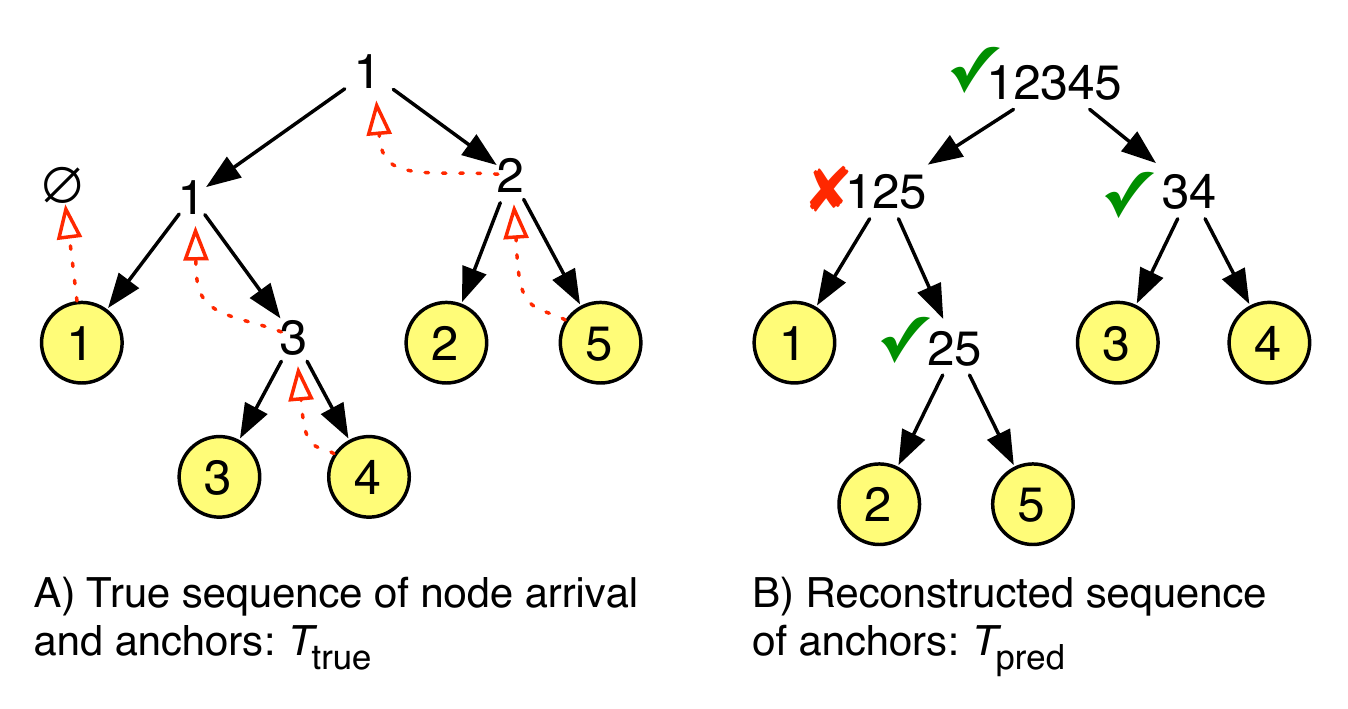}
\caption{\label{fig:rfdist}Computing the similarity of node/anchor pairs in the
true versus the reconstructed histories.}
\end{figure}

When a node enters the network under the DMC and FF models, it chooses an
existing node from which it copies links. We call this node its \emph{anchor}.
To assess our ability to identify node/anchor relationships, we encode the true
node/anchor relationships in a binary tree. We can think of a node's arrival as
causing its chosen anchor node to divide in two, producing a new node and a new
copy of the old node. Figure~\ref{fig:rfdist}A shows a binary tree describing
such a bifurcation process, with node anchors indicated by dotted arrows. In
this example, node~$1$ initially exists alone in the network, and therefore has
no anchor.  Reading from top down, node~$2$ enters and chooses node $1$ as its
anchor. This spawns a new node~$1$, which is conceptually different from its
parent because the new node could have gained or lost edges due to the arrival
of node~$2$.  Node~$3$ enters and chooses the new node~$1$ as its anchor.
Finally, nodes~$4$ and~$5$ anchor from nodes~$3$ and~$2$, respectively.

Figure~\ref{fig:rfdist}B shows an example sequence of merges predicted by our
reconstruction algorithms.  Internal nodes in the tree are labeled with the
concatenation of the labels of its two children indicating an inferred
node/anchor relationship between the children.

Let $T_{\textrm{true}}$ be the anchor tree derived from the true growth process
(Figure~\ref{fig:rfdist}A) and let $T_{\textrm{pred}}$ be the reconstructed
anchor tree (Figure~\ref{fig:rfdist}B).  We can assess the quality of the
reconstruction by seeing what percentage of subtrees in $T_{\textrm{pred}}$ are
found in $T_{\textrm{true}}$.  This measure (called \textbf{Anchor}) is closely
related to the Robinson-Foulds distance metric used to compare phylogenetic
trees~\cite{Felsenstein2003}. In the example of Figure~\ref{fig:rfdist}, the
similarity between the trees is $3/4 = 75\%$.

This validation measure is advantageous because it evaluates if the
relationship between larger groups of nodes was correctly determined. In
addition, it does not unduly penalize the mis-ordering of arrival times for
nodes that are far apart in the network.  It also does not depend on which node of the
merged pair $(u,v)$ was deleted from the graph in the DMC model,
because both choices lead to the same subtree in $T_\textrm{pred}$.  On the
other hand, the measure is in some ways stricter than counting correct
node/anchor pairs. For example, in Figure~\ref{fig:rfdist} it would be
incorrect to merge $1$ and $2$ in the first backward step because the extant
nodes $1$ and $2$ are not the same as the past nodes $1$ and $2$.

%==========================================================================
%==========================================================================
\section{Results}

%==========================================================================
\begin{raggedright}
\subsection{Model reversibility using the greedy likelihood algorithm}
\end{raggedright}

\begin{figure*}[t]
\centering
\includegraphics[width=\textwidth]{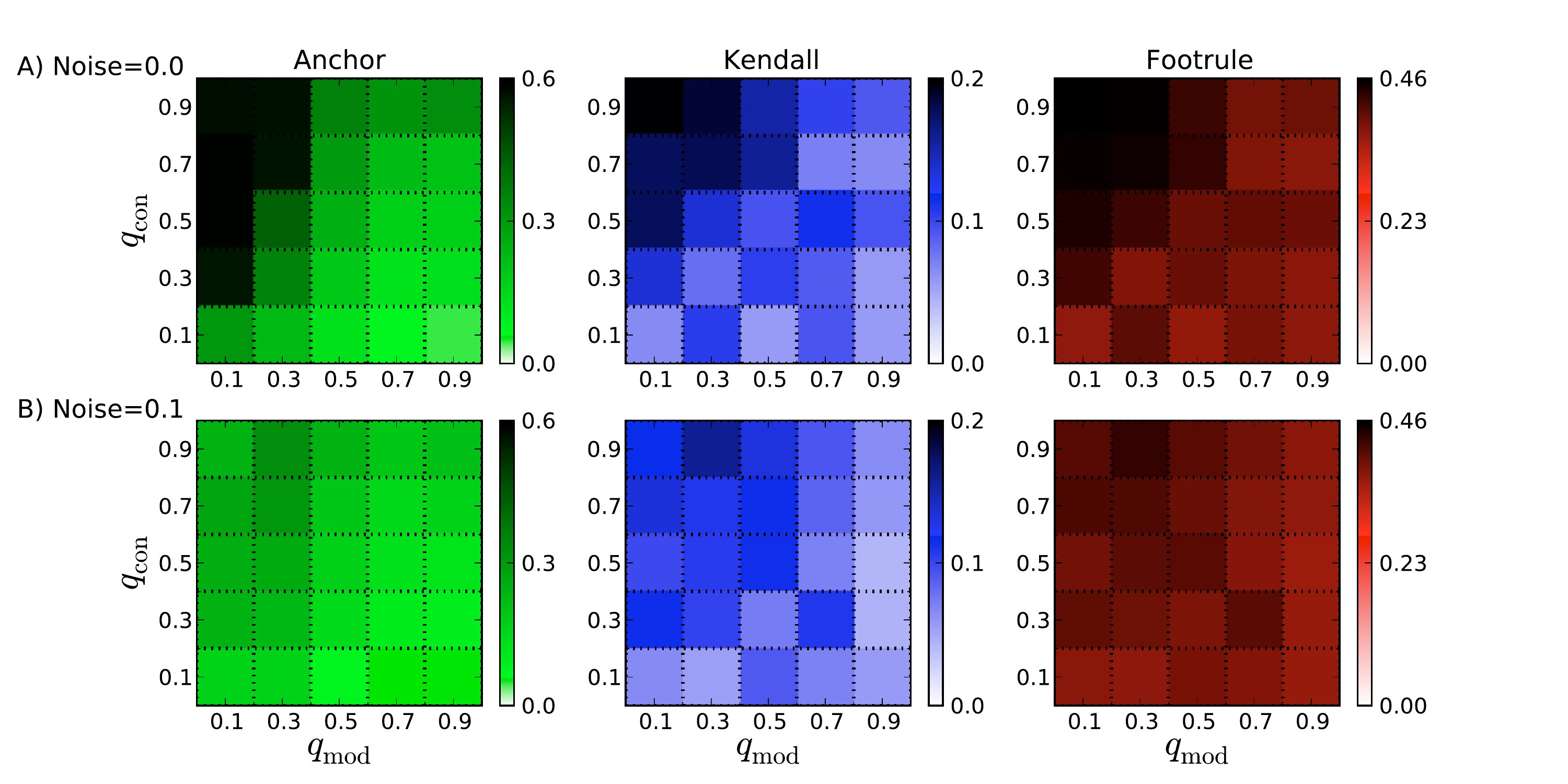}
\caption{\label{fig:dmc}Accuracy of node arrival times and node anchors using
the DMC model. The $x$ and $y$-axes show the DMC parameters ($\qmod,\qcon$) used to grow the synthetic network forward. Each parameter varies from $0.1$--$0.9$ in steps of $0.2$.
The intensity of each cell in the heatmap represents the quality of the reconstruction validation measure (Anchor, Kendall, Footrule) under optimal reverse parameters. (A) and (B) show results under varying levels of noise. For many DMC-grown synthetic networks, accurate reconstruction is possible.}
\end{figure*}

We first tested the algorithms in situations where the evolutionary history is
completely known.  This allows us to assess the performance of the greedy
likelihood algorithm and to compare the reversibility of various network
models.  For each model (and choice of parameters), we grew $100$-node networks
forward according to the model, and then supplied only the final network
$G_{t=100}$ to our algorithm to reconstruct its history.  We repeated this
process $10$ times and averaged the results for each combination.

For the DMC model under realistic choices of $\qmod$ and $\qcon$, almost $60\%$
of the node/anchor relationships inferred are correct if the optimal choice of
$\qmod$ and $\qcon$ parameters are used in the reconstruction process.
Figure~\ref{fig:dmc}A plots the performance of the $3$ validation measures for
$25$ combinations of $(\qmod,\qcon)$ model parameters.  Both the Spearman's
footrule and Kendall's $\tau$ measures of arrival-time correlation reveal an ability
to order nodes correctly significantly better than random.  In general, it is
difficult to predict all arrival times correctly because unrelated duplications
could occur in successive steps in completely different parts of the graph. But
still, a large agreement can be obtained from analysis of the final graph
alone.

Reversibility varies drastically depending on the DMC model parameters used to
grow the network forward.  Naturally, increasing $\qmod$ induces more random
changes in the network, which makes it more difficult to reverse the evolution.
Conversely, as $\qcon$ increases, the history generally becomes easier to
reverse because more nodes are directly connected to the node from which they
duplicated.

\begin{figure*}[t]
\centering
\includegraphics[width=1.0\textwidth]{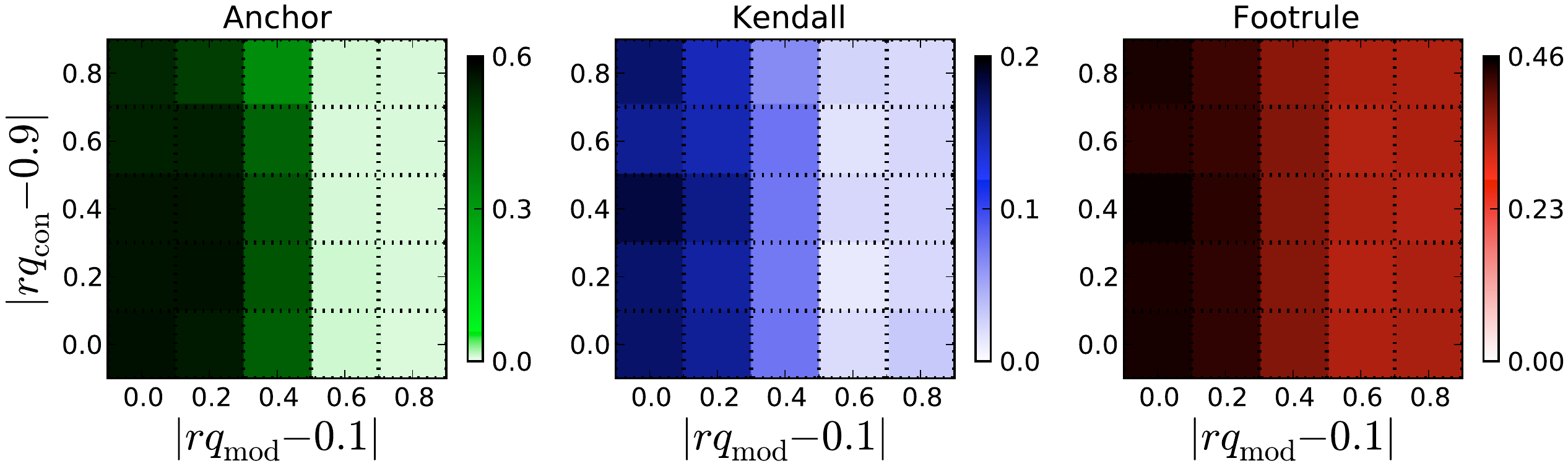}
\caption{\label{fig:dmcback}Accuracy of node arrival times and node anchors when reverse parameters are not known. Synthetic DMC-grown networks were constructed using $\qmod=0.1,\qcon=0.9$ and reversed using all $25$ combinations of reversal parameters. The $x$ and $y$-axes show the difference between the reversal parameters ($r\qmod$ and $r\qcon$, respectively) and the forward parameters ($0.1$ and $0.9$, respectively). The intensity of each cell in the heatmap represents the quality of the reconstruction validation measure (Anchor, Kendall, Footrule). Accurate histories can be inferred as long as reverse parameters (in particular, $r\qmod$) are in the rough range of the forward parameters.}
\end{figure*}

Performance also depends on the match between the values of $\qmod$ and $\qcon$
used to grow the network forward and those used to reverse the history
(Figure~\ref{fig:dmcback}).  However, even if the forward parameters are not
known exactly, it is feasible to reconstruct a meaningful history if the
reversal parameters are chosen to be near the forward parameters. There is
often a hard transition at $\qmod=0.5$ or $\qcon=0.5$ when the bias towards
having an edge and not having an edge tips to one side or the other.  Though
optimal performance can correspond to reversing a network with the same
parameters used to grow the network, this need not be the case. For example,
suppose $30$\% of all nodes have edges to their anchors. This does not imply
that setting $\qcon = 0.3$ will work best because the true pair sought will
likely not be connected and hence even lower values of $\qcon$ may lead to a
more accurate reconstruction.

\begin{figure}[t]
\centering
\includegraphics[width=\columnwidth]{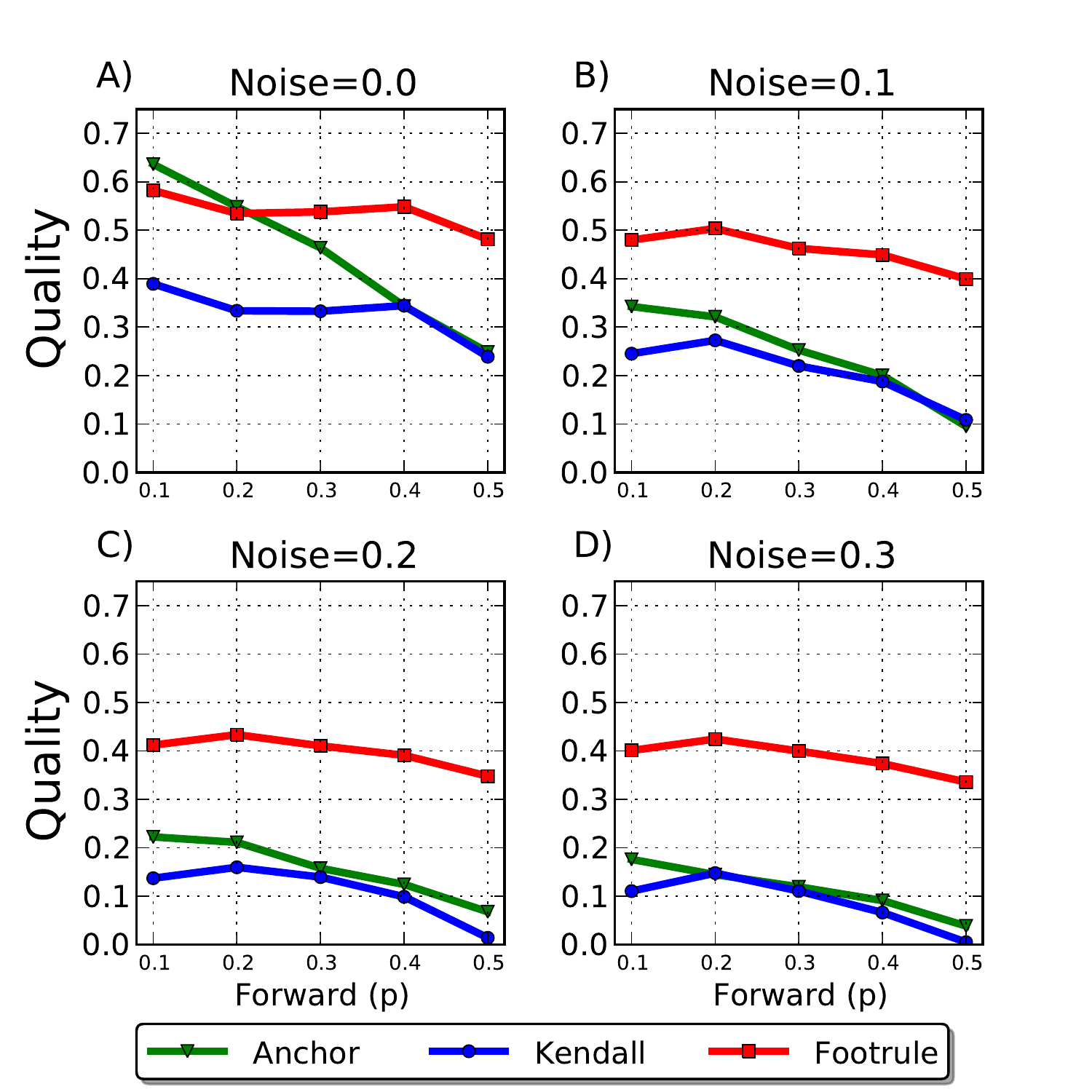}
\caption{\label{fig:ff}Accuracy of arrival times and node anchors using the
forest fire model.  (A--D) The $x$-axis shows the FF parameter ($p$) used to
grow the synthetic network forward. The $y$-axis shows the quality of the $3$
reconstruction validation measures under optimal reverse parameters. All
FF-based reconstructions are significantly better than random reconstructions,
even when $30\%$ of true edges are replaced by random edges.}
\end{figure}

% Forest fire

We performed the same synthetic-data experiments using the forest fire model
for varying values of the parameter $p$, which controls the spread of the fire,
ranging from $0.1$ to $0.5$. (Values of parameter $p>0.5$ resulted in mostly
clique-like networks.) Figure~\ref{fig:ff}A shows that between $25\%$ and
$64\%$ of anchor relationships can be correctly identified, and that the
estimated node arrival ordering resembles the true arrival order.  As
$p$ increases, performance of all measures tends to decrease. This is because
as $p$ increases, the degree of each node increases, thus making it more
difficult to pick out the correct anchor from among the set of neighbors.

\begin{figure}[t]
\centering
\includegraphics[width=\columnwidth]{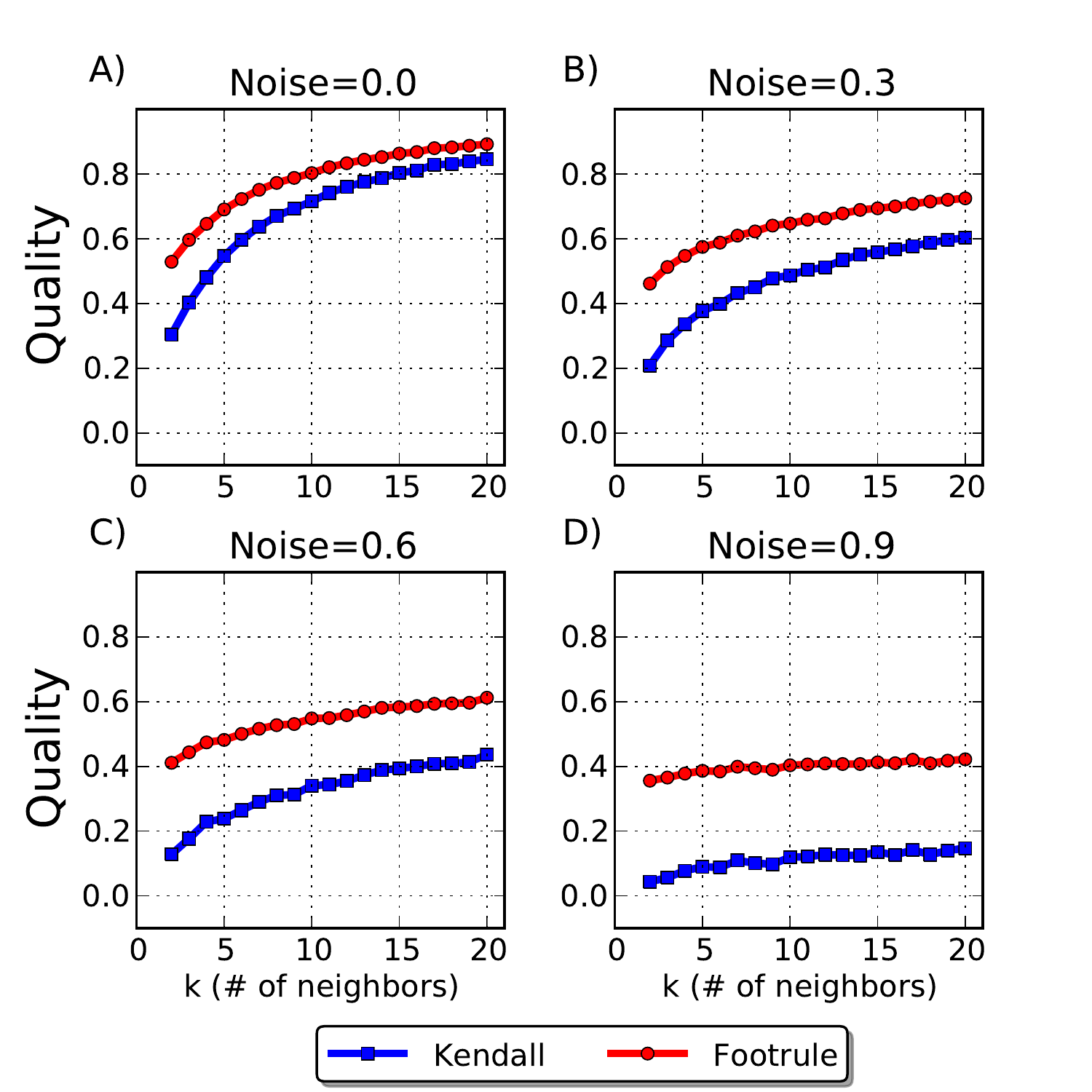}
\caption{\label{fig:bpa}Agreement with arrival times using the preferential
attachment model.  (A--D) The $x$-axis shows the PA parameter ($k$) used to grow
the synthetic network forward. The $y$-axis show the quality of the $3$
reconstruction validation measures. Compared to the DMC and FF models, the PA
model is easiest to reverse, even in the presence of substantial noise.}
\end{figure}

% Pref attachment

Finally, we grew $1000$-node networks using the linear preferential attachment
model for various choices of parameter $k$, the number of neighbors to which a new node
initially connects (Figure~\ref{fig:bpa}).  Of the three models we consider, PA
is the most easily reversible.  As $k$ increases, there becomes more
opportunity for older and newer nodes to differentiate themselves from one
another, and hence the network becomes easier to reverse.
Figure~\ref{fig:bpa}A shows that for the PA model we can achieve Kendall $\tau$
values of over $80$ percentage points higher than random when $k>15$.  In the
PA model, a new node does not choose an anchor node to copy links from so only
the arrival-time validation measures are applicable.

%==========================================================================
\begin{raggedright}
\subsection{Effect of deviation from the assumed model}
\vspace{-5pt}
\end{raggedright}

To gauge robustness to deviations from the growth model, we repeated the
experiments on synthetic data after randomly replacing some percentage of edges
in the final graph with new edges.  Under all models, reconstruction quality
generally suffers from a noisy view of the present-day graph but meaningful
histories can still be recovered.

DMC is the most sensitive to the addition of noise (Figure~\ref{fig:dmc}B),
while PA is by far the most resilient to noise. Even when $90\%$ of the true
edges are replaced with random edges, nearly turning the graph into a random
graph, reversibility of PA is still better than random (Figure~\ref{fig:bpa}D).
DMC can tolerate noise up to $30\%$ before returning essentially random
reconstructions. The robustness of the forest fire model lies in between DMC
and PA (Figure~\ref{fig:ff}D).

Mis-identifying the model used to grow the network can also significantly
reduce the quality of the inferred history. To verify this, we grew networks
forward using DMC $(\qmod=0.1, \qcon=0.9)$ and reversed it with the other
models. The low $\qmod$ value implies that a node has many reasonable anchors.
A reversal using the FF model cannot distinguish between these many reasonable
anchors.  In particular,  FF performs approximately 10 times
worse than DMC according to both the Spearman's footrule and Kendall's $\tau$ measures.
Further, FF is only able to uncover an average of $4\%$ of correct node/anchor
relationships compared to $55\%$ using DMC. PA also performs poorly in this
case because nodes with late arrival times can duplicate from hubs and
immediately become ``hubby''.
Hence, reversing DMC-grown networks involves more than
removing low-degree nodes.  As $\qmod$ increases, FF and PA each perform better
at reversing DMC networks, but both still perform worse than DMC (e.g.\@ at
$\qmod=0.5$, FF and PA have average Kendall $\tau$ values of $9\%$ and $10\%$,
respectively, compared to $15\%$ for DMC).

Random networks grown forward using PA are best reversed using PA as opposed to
DMC or FF. At $k=10$, PA has a Kendall $\tau$ value of over $70\%$ compared to
only $36\%$ for DMC and $20\%$ for FF. At higher values of $k$, this difference
is even more pronounced. The reason DMC and FF perform so poorly is because,
for each node, they seek a single anchor from which the observed
links can be explained. With PA, however, a node can have neighbors that are
far apart in the network.

%==========================================================================
\begin{raggedright}
\subsection{Recovery of ancient protein interaction networks}
\end{raggedright}

\begin{figure}[t]
\centering
\includegraphics[width=\columnwidth]{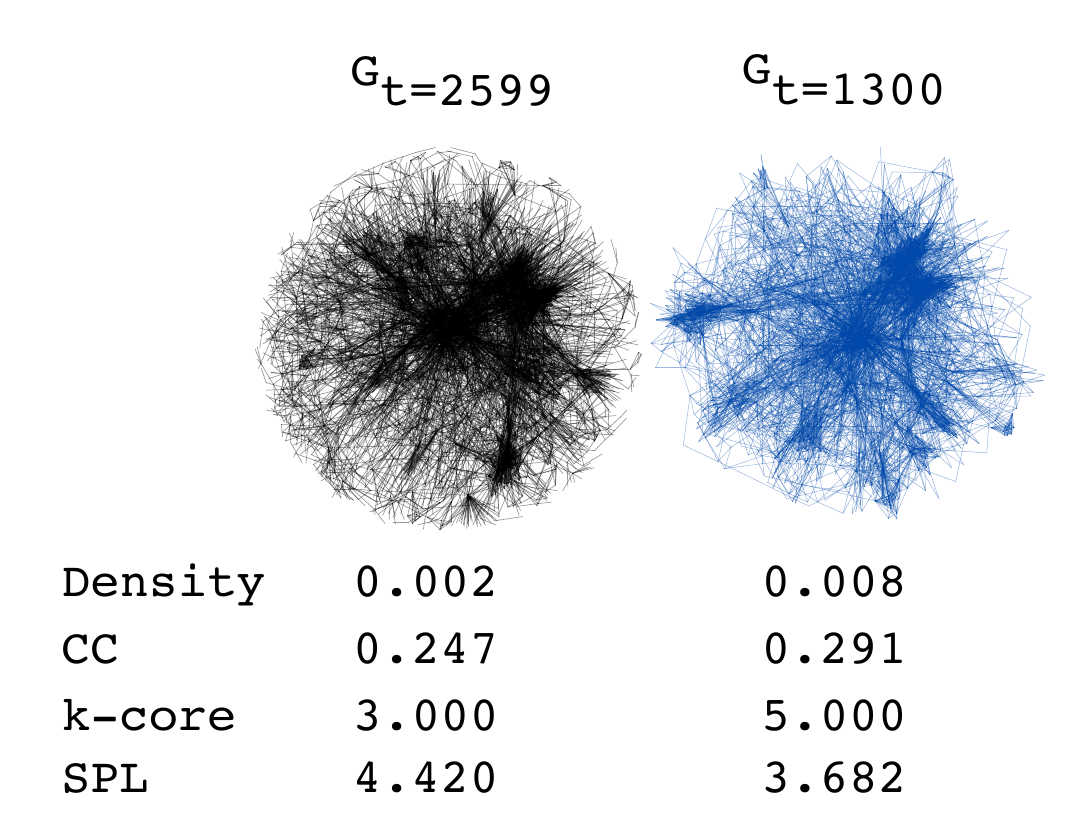}
\caption{\label{fig:netviz} Visualization~\cite{Cytoscape} of the extant PPI network ($G_{t=2599}$) and an ancestral version ($G_{t=1300}$). The density, clustering coefficient (CC), average shortest path length (SPL), and average $k$-core number are shown for each network. The ancient network is considerably denser than the extant network.}
\end{figure}

We obtained a high-confidence protein-protein interaction (PPI) network for the
yeast \emph{S.\@~cerevisiae} from the IntAct database~\cite{Kerrien2007}. The
network contains $2,599$ proteins (nodes) and $8,275$ physical interactions
between them. We applied the reversal algorithm for $2,599$ steps to estimate a
complete history of the growth of the network. Figure~\ref{fig:netviz} shows the
original network ($G_{t=2599}$) and an inferred ancestral network with $1300$
nodes ($G_{t=1300}$).

Because PPI networks from the past are unavailable, we do not directly have
true node arrival times to which we can compare.  Instead, we estimate protein
arrival times using sequence-based homology under the assumption that proteins
that have emerged after yeast diverged from other species will have fewer
orthologs in these distantly related organisms~\cite{Li2004}. In particular, we
obtained data for the occurrence of orthologs of yeast proteins in 6 eukaryotes
(\emph{A.\@~thaliana}, \emph{C.\@~elegans}, \emph{D.\@~melanogaster},
\emph{H.\@~sapiens}, \emph{S.\@~pombe}, and \emph{E.\@~cuniculi}) from the
Clusters of Orthologous Genes database~\cite{Tatusov2003}.  The number of
species for which an ortholog was present was used as a proxy for the arrival
time: proteins with orthologs in $6$ of the eukaryotes are likely to be older
than proteins with orthologs in~$5$, and so on.

\begin{figure*}[t]
\centerline{\includegraphics[width=0.65\textwidth]{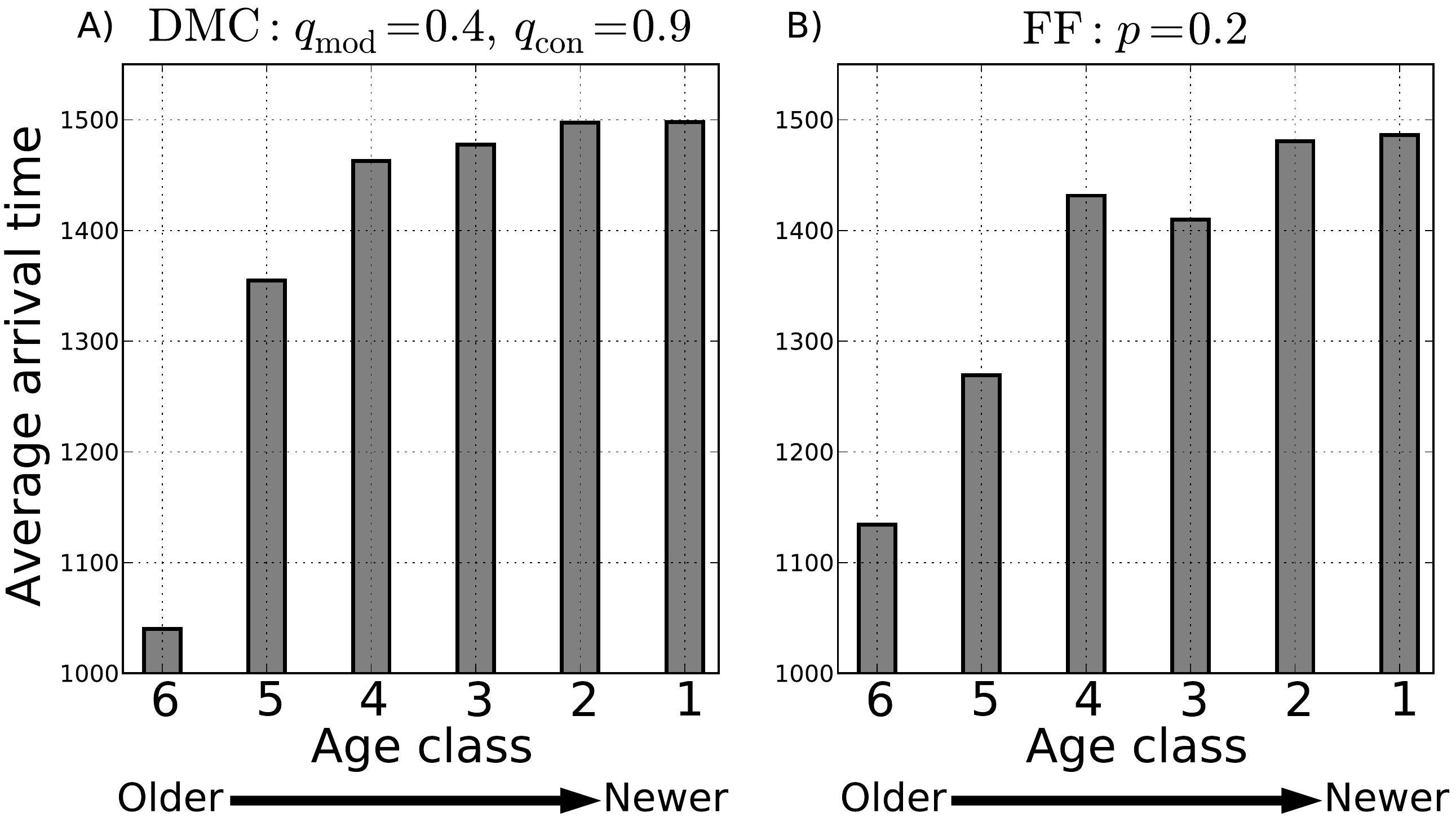}}
\caption{\label{fig:intact2}Predicting protein age groups by reversing the DMC
and FF models on a real PPI network for \emph{S.\@~cerevisiae}. The $x$-axis
shows the 6 age classes for proteins. The $y$-axis shows the average arrival
time for proteins in the class. The DMC model correctly orders all classes, whereas the FF model swaps classes 3 and 4.}
\end{figure*}

Reversing the network using the DMC model produced an estimated node arrival
order in greater concordance with the orthology-based estimates of protein age
than either the FF or PA models.  Figure~\ref{fig:intact2} shows the average
reconstructed arrival time for proteins in each of the $6$ age classes for the
DMC and FF models (the results for PA were similar to FF).  The results shown
are the best for each model over the tested parameter space and thus represent
the limit of performance for each of the models using the proposed algorithm.
The DMC model (Figure~\ref{fig:intact2}A) correctly determines the relative
ordering of all the age classes ($P$-value $<0.01$ after Bonferroni correcting
for optimal parameter usage). While the FF model does reconstruct the ordering
of some of the age classes, it never produces the exact ordering for any choice
of parameters (consistently swapping classes $3$ and $4$;
Figure~\ref{fig:intact2}B). This provides additional
evidence~\cite{Middendorf2005} that a DMC-like model is a better fit for PPI
networks than models such as FF and PA inspired by social networks.

%==========================================================================
\begin{raggedright}
\subsection{Estimation of parameters governing network growth}
\end{raggedright}

The parameters that produced the history that best matched the sequence-based estimates of protein ages
provide hints about the relative importance of various
processes in network growth. For DMC, the optimal parameters were $\qmod=0.4$
and $\qcon=0.9$.  We can use these as estimates of the probability that an
interaction is modified following a gene duplication ($\approx50$\%) and the
probability that two duplicated genes interact (high, as also found
elsewhere~\cite{Ispolatov2005HD,Pereira-Leal2007,NavlakhaJCB2009}).

Interestingly, the optimal FF and DMC parameters create models that have many
similarities.  Optimal performance was obtained for the FF model with parameter
$p = 0.2$, which implies that both the anchor and the arriving node will have
similar neighborhoods because the simulated fire likely does not spread beyond
the immediate neighbors of the anchor. The property of similar neighborhoods is
also implied by duplication step of DMC coupled with the moderate value of
$\qmod=0.4$. Further, in the FF model the arriving node is always linked to its
anchor, and the high value of $\qcon=0.9$ causes this to frequently happen in
the DMC model as well.  Thus, based on their agreement with sequence-based
estimates of protein arrival times, two independent and very different base
models both suggest that proteins should very frequently interact with the
protein from which they duplicated, and that the new node should primarily
interact with neighbors of their anchors.

%==========================================================================
\begin{raggedright}
\subsection{Protein complexes and evolution by duplication}
\end{raggedright}

We can test correctness of the anchors identified by DMC and FF using
protein annotations. A protein and its duplicate are often involved in similar
protein complexes in the cell~\cite{Pereira-Leal2006,Pereira-Leal2007}.  We
expect then that the node/anchor pairs identified ought to correspond to
proteins that are co-complexed.  Because it is difficult to model the gain and
loss of functional properties of ancient proteins, we only tested this
hypothesis among pairs of extant proteins.

Using the MIPS complex catalog~\cite{Guldener2007}, which contained annotations
for $994$ of the proteins in the network, $79\%$ of the testable node/anchor
pairs predicted using the DMC model shared an annotation. This is much higher
than the baseline frequency: only $55\%$ of edges in the extant network connect
nodes that share an annotation. Under the FF model, $68\%$ of node/anchor pairs
share a MIPS annotation. So, while the FF model under this validation measure
again is performing much better than expected by random chance, it does not
perform as well as DMC. The high quality of the DMC-based node/anchor pairs
also supports the idea that a good definition of a functional module in a PPI
network is one which groups proteins with similar neighbors together (rather
than one based strictly on density)~\cite{NavlakhaJCB2009}.

\begin{figure*}[t]
\centering
\includegraphics[width=0.48\textwidth]{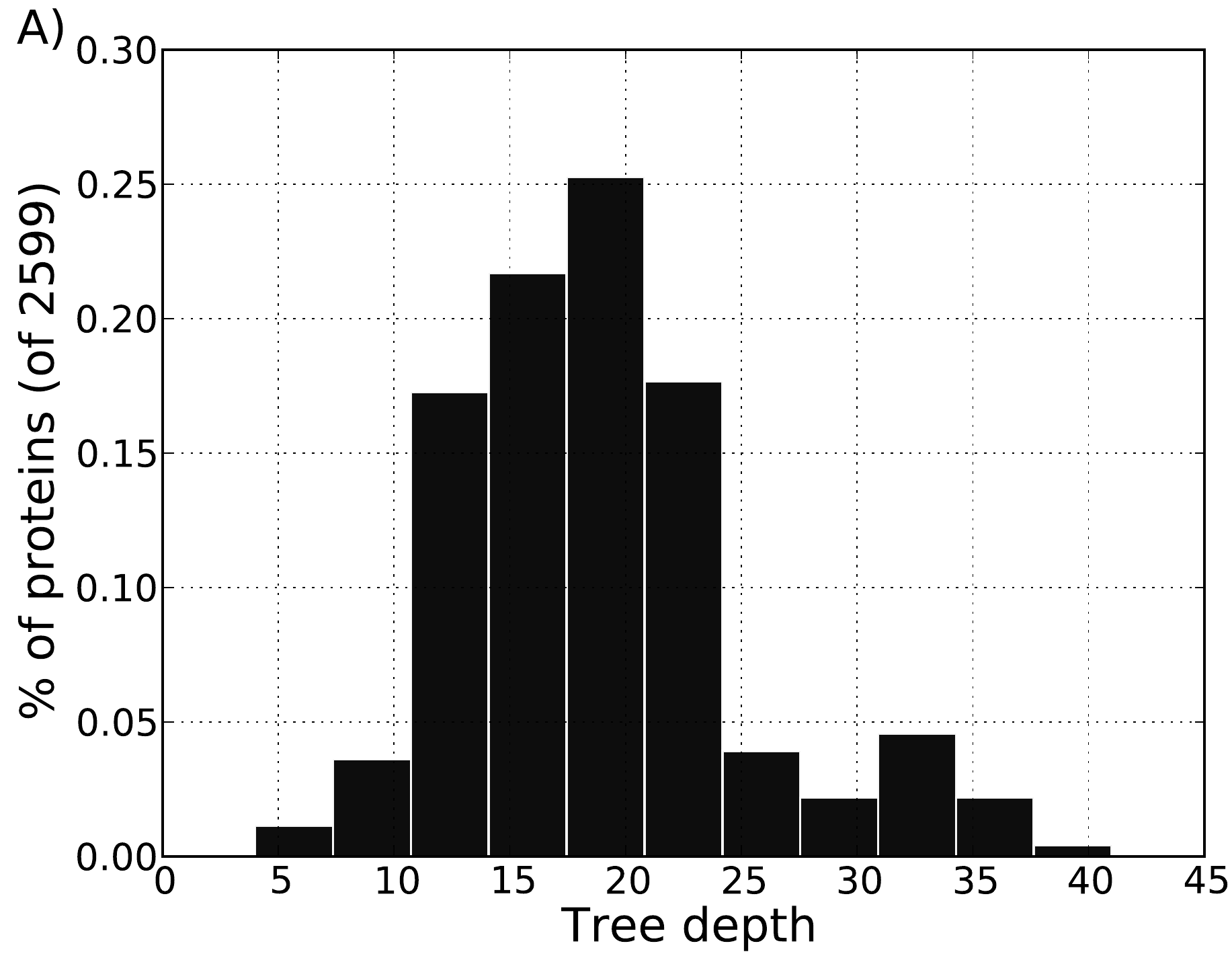}
\includegraphics[width=0.48\textwidth]{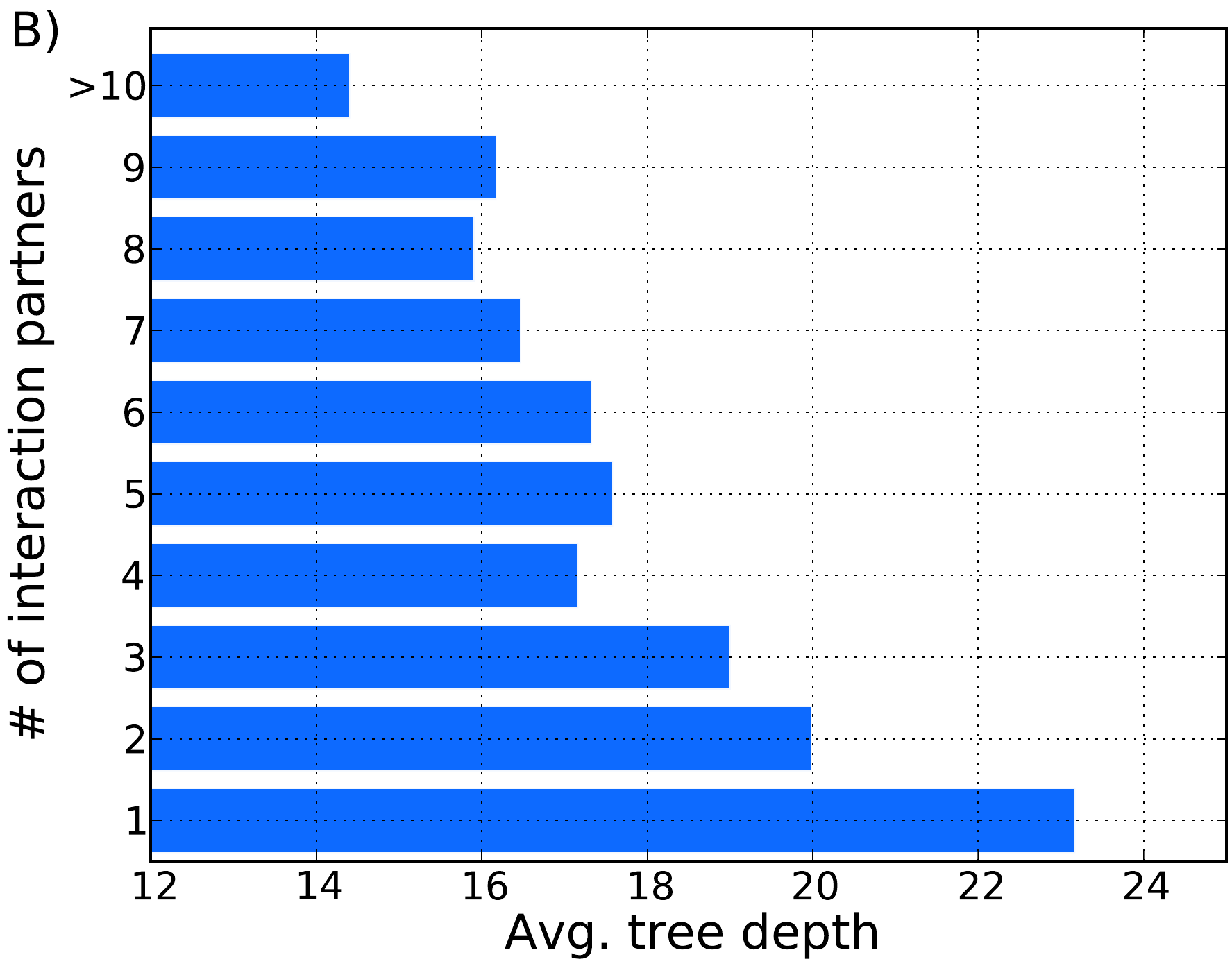}
\caption{\label{fig:treedepth} (A) The distribution of duplication rates for extant proteins in the PPI network. The $x$-axis of the histogram is the number of duplications, measured as the distance from the root of the phylogeny to the extant protein. The $y$-axis is the percentage of proteins lying in the tree depth bin. (B) The relationship between duplication and number of interaction partners. The $x$-axis shows the average tree depth for proteins with the given number of interaction partners ($y$-axis) in $G_{t=2599}$.}
\end{figure*}

% tree depth.

The phylogeny of node/anchor relationships
can also help characterize how duplication has guided the evolution of the
yeast proteome.  We estimate the number of times each extant protein was
involved in a duplication (that becomes fixed in the population) by computing the
depth of the protein in the inferred node/anchor tree.  Figure~\ref{fig:treedepth}A
shows that most proteins are involved in a similar number of duplications (mean
= 19), with fewer proteins involved in many more or many less.
Proteins involved in more duplications typically have fewer interaction
partners (Figure~\ref{fig:treedepth}B). Using network histories alone, this
confirms previous sequence-based findings that the evolutionary rate of
proteins is inversely proportional to its number of binding
partners~\cite{Fraser2002,Makino2006}.

% protein complex evolution.

The arrival times of proteins can also tell us how different components of
protein complexes might have evolved. For every protein belonging to exactly
one MIPS complex, we computed its \emph{coreness}, defined as the percentage of its
annotated neighbors that belong to the same complex. A large coreness value
indicates that the protein plays a central role in the complex; a small value
suggests a peripheral role~\cite{Gavin2006}. Amongst the $763$ protein tested,
there was a significant correlation between older proteins and larger coreness
values ($R=0.36$,~$P$-value~$<0.01$), a trend that \citet{Kim2008} also
independently reported by studying the evolution of protein structure using a
different measure of coreness.

%==========================================================================
\subsection{Recovery of past social networks}

To contrast the evolution of biological networks with social networks, we
applied our algorithms to part of the Last.fm music social network.  Edges in
this network link users (nodes) that are friends. We sampled a region of the  network by performing a
breadth-first crawl starting from a random user `\texttt{rj}'. We recorded the
date and time of registration for each node visited, which corresponds to its
arrival time into the network.  The resulting network consisted of the subgraph
induced by the first $2,957$ nodes visited ($9,659$ edges). Because only a
subgraph of the complete network was visited, some nodes have neighbors that
are outside the induced subgraph.  This missing data makes the reconstruction
problem even more difficult.

\begin{figure}[t!]
\centerline{\includegraphics[width=\columnwidth]{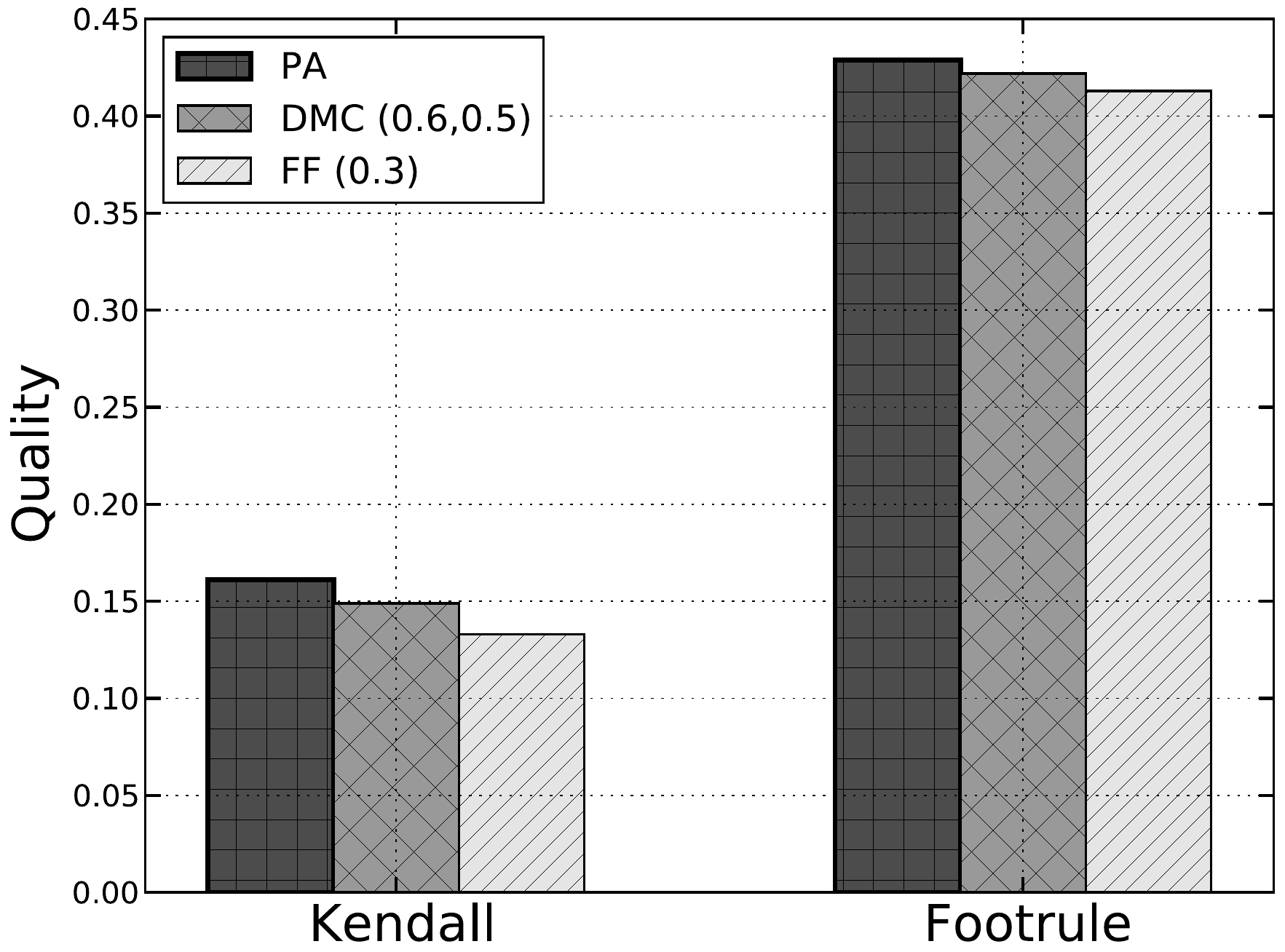}}
\caption{\label{fig:lastfm}Predicting node arrival times for users in the
Last.fm social network. The PA model appears most applicable to reversing the
network.}
\end{figure}

Figure~\ref{fig:lastfm} shows the performance of the models (using the best
parameters) for the node-arrival measures. The best performing model
(preferential attachment) for the Last.fm network was the worse performing
model for the PPI network.  Further, the optimal FF value of $p$ for the
Last.fm network was larger ($0.3$) than for the PPI network ($0.2$). This fits
well with the notion that new users in social networks often form links to a
varied set of existing users that might be far apart in the
network~\cite{Leskovec2005}.

An advantage of FF and DMC over PA is that the former return node
anchors. To validate these predicted relationships, we make the observation
that node/anchor pairs are likely to share similar taste in music. As a null
baseline, we computed the percentage of edges in the given network $G_{t=2957}$
that connect users who share a top-5 favorite artist. The pairs returned by FF
are significantly more likely ($13.8\%$) to share a top-5 favorite artist over
DMC $(10.3\%)$ and the baseline $(10.8\%)$.  Most users act as anchors to $\leq
1$ new member, however, there were $9$ users who (putatively) each brought
$\geq 30$ new members into the network. Such popular anchors can be thought of
as members who are responsible for the network's organic growth.

%==========================================================================
%==========================================================================
\section{Discussion}

We presented a novel framework for uncovering precursor versions of a network
given only a growth model by which the network putatively evolved.  Our
approach works backwards from a given network and is therefore network specific
(not model generic) and can retain individual node labels.

Using the proposed algorithms, we estimated protein ages from the topology of a
PPI alone that matched sequence-based estimates well.  Further, we correlated
node/anchor pairs with co-complexed proteins and characterized the distribution
of duplications on a per-protein basis. We also found that older proteins tend to
play a more central role in protein complexes than newer, peripheral proteins.
Given the noisy and incomplete status of the available PPI data and the simple
network growth models, it is surprising that such high agreement with known
biology can be obtained.

We also used the accuracy of history reconstruction as an optimization
criterion for choosing model parameters. We determined, via both the DMC and FF
models, that duplicated proteins are likely to interact and share many
interaction partners. The ability to match the inferred history under a given
model to properties of the true history provides an alternative way to validate
models that goes beyond comparing only statistics of the final extant network.

A heuristic approach that orders node arrival times based on their static
degree in the extant network performs similarly to PA (poor performance on
DMC-grown synthetic networks and the PPI network; better performance for
PA-grown synthetic networks and the Last.fm network). PA is derived from the
assumption that degree distribution is correlated with age, hence the
similarity is to be expected. The drawback with such a heuristic is
that it does not provide a likelihood estimate for a history, in contrast to our more
principled approach.

A natural extension to this work is to evaluate how the greedy likelihood
approach performs on other models, such as those that explicitly incorporate an
estimate of a node's age~\cite{Middendorf2005,Kim2008} or those in which nodes
can add edges at variable times~\cite{LeskovecKDD2008}.  Automated selection of
reverse model parameters and computation of model-based priors to use in the
likelihood procedure may make the reconstructions more accurate and more
practical. However, even with the standard models investigated here, our
results show that present-day networks are strongly linked to their past, and
that this past can be effectively excavated.

%==========================================================================
%==========================================================================
\section*{Acknowledgements}

The authors thank Geet Duggal, Justin Malin, Guillaume Mar\c{c}ais, and Galileo
Namata for helpful conversations about the manuscript. C.K. thanks the National
Science Foundation for grants 0849899 and 0812111.

\bibliographystyle{apalike}
\bibliography{timemachine}

\begin{thebibliography}{}

\bibitem[Ahmed and Xing, 2009]{Ahmed09}
Ahmed, A. and Xing, E.~P. (2009).
\newblock Recovering time-varying networks of dependencies in social and
  biological studies.
\newblock {\em Proc.\@ Natl.\@ Acad.\@ Sci.\@ USA}, 106:11878--11883.

\bibitem[Bar-Ilan et~al., 2006]{Bar-Ilan2006}
Bar-Ilan, J., Mat-Hassan, M., and Levene, M. (2006).
\newblock Methods for comparing rankings of search engine results.
\newblock {\em Comput. Netw.}, 50(10):1448--1463.

\bibitem[Barab\a'asi and Albert, 1999]{Barabasi1999}
Barab\a'asi, A.~L. and Albert, R. (1999).
\newblock Emergence of scaling in random networks.
\newblock {\em Science}, 286(5439):509--512.

\bibitem[Bez\'akov\'a et~al., 2006]{Bezakova06}
Bez\'akov\'a, I., Kalai, A., and Santhanam, R. (2006).
\newblock Graph model selection using maximum likelihood.
\newblock In {\em Proc.\@ 23rd Intl.\@ Conf.\@ on Maching Learning}, pages
  105--112.

\bibitem[Dutkowski and Tiuryn, 2007]{Dutkowski2007}
Dutkowski, J. and Tiuryn, J. (2007).
\newblock Identification of functional modules from conserved ancestral
  protein-protein interactions.
\newblock {\em Bioinformatics}, 23(13):i149--i158.

\bibitem[Felsenstein, 2003]{Felsenstein2003}
Felsenstein, J. (2003).
\newblock {\em Inferring Phylogenies}.
\newblock Sinauer Associates, 2nd edition.

\bibitem[Flannick et~al., 2006]{Flannick2006}
Flannick, J., Novak, A., Srinivasan, B.~S., McAdams, H.~H., and Batzoglou, S.
  (2006).
\newblock Graemlin: general and robust alignment of multiple large interaction
  networks.
\newblock {\em Genome Res.\@}, 16(9):1169--1181.

\bibitem[Fraser et~al., 2002]{Fraser2002}
Fraser, H.~B., Hirsh, A.~E., Steinmetz, L.~M., Scharfe, C., and Feldman, M.~W.
  (2002).
\newblock Evolutionary rate in the protein interaction network.
\newblock {\em Science}, 296(5568):750--752.

\bibitem[Gavin et~al., 2006]{Gavin2006}
Gavin, A.-C., Aloy, P., Grandi, P., Krause, R., Boesche, M., Marzioch, M., Rau,
  C., Jensen, L.~J., Bastuck, S., D\"{u}mpelfeld, B., Edelmann, A., Heurtier,
  M.-A., Hoffman, V., Hoefert, C., Klein, K., Hudak, M., Michon, A.-M.,
  Schelder, M., Schirle, M., Remor, M., Rudi, T., Hooper, S., Bauer, A.,
  Bouwmeester, T., Casari, G., Drewes, G., Neubauer, G., Rick, J.~M., Kuster,
  B., Bork, P., Russell, R.~B., and Superti-Furga, G. (2006).
\newblock Proteome survey reveals modularity of the yeast cell machinery.
\newblock {\em Nature}, 440:631--636.

\bibitem[Gibson and Goldberg, 2009]{Gibson2009}
Gibson, T.~A. and Goldberg, D.~S. (2009).
\newblock Reverse engineering the evolution of protein interaction networks.
\newblock {\em Pac. Symp. Biocomput.}, pages 190--202.

\bibitem[Golbeck, 2007]{Golbeck2007}
Golbeck, J. (2007).
\newblock The dynamics of web-based social networks: Membership, relationships,
  and change.
\newblock {\em First Monday}, 12(11).

\bibitem[Guldener et~al., 2005]{Guldener2007}
Guldener, U. et~al. (2005).
\newblock {CYGD}: the comprehensive yeast genome database.
\newblock {\em Nucleic Acids Res.}, 33(Suppl.\@ 1):D364+.

\bibitem[Guo et~al., 2007]{Guo07}
Guo, F., Hanneke, S., Fu, W., and Xing, E. (2007).
\newblock Recovering temporally rewiring networks: A model-based approach.
\newblock In {\em Proc.\@ 24th Intl.\@ Conf.\@ on Machine Learning}, pages
  321--328.

\bibitem[Hanneke and Xing, 2007]{Hanneke06}
Hanneke, S. and Xing, E. (2007).
\newblock Discrete temporal models of social networks.
\newblock In {\em Proc.\@ 23rd Intl.\@ Conf.\@ on Maching Learning, Workshop on
  Statistical Network Analysis}, pages 115--125.

\bibitem[Hopcroft et~al., 2004]{Hopcroft2004}
Hopcroft, J., Khan, O., Kulis, B., and Selman, B. (2004).
\newblock Tracking evolving communities in large linked networks.
\newblock {\em Proc.\@ Natl.\@ Acad.\@ Sci.\@ USA}, 101(Suppl.\@ 1):5249--5253.

\bibitem[Hormozdiari et~al., 2007]{Hormozdiari07}
Hormozdiari, F., Berenbrink, P., Pr\u{z}ulj, N., and Sahinalp, S.~C. (2007).
\newblock Not all scale-free networks are born equal: The role of the seed
  graph in {PPI} network evolution.
\newblock {\em PLoS Comput.\@ Biol.\@}, 3(7):e118.

\bibitem[Ispolatov et~al., 2005a]{Ispolatov2005}
Ispolatov, I., Krapivsky, P.~L., and Yuryev, A. (2005a).
\newblock Duplication-divergence model of protein interaction network.
\newblock {\em Phys Rev. E}, 71(6 Pt 1):061911.

\bibitem[Ispolatov et~al., 2005b]{Ispolatov2005HD}
Ispolatov, I., Yuryev, A., Mazo, I., and Maslov, S. (2005b).
\newblock Binding properties and evolution of homodimers in protein-protein
  interaction networks.
\newblock {\em Nucleic Acids Res.\@}, 33(11):3629--3635.

\bibitem[Kelley et~al., 2003]{Kelley2003}
Kelley, B.~P., Sharan, R., Karp, R.~M., Sittler, T., Root, D.~E., Stockwell,
  B.~R., and Ideker, T. (2003).
\newblock Conserved pathways within bacteria and yeast as revealed by global
  protein network alignment.
\newblock {\em Proc.\@ Natl.\@ Acad.\@ Sci.\@ USA}, 100(20):11394--11399.

\bibitem[Kerrien et~al., 2007]{Kerrien2007}
Kerrien, S. et~al. (2007).
\newblock Intact--open source resource for molecular interaction data.
\newblock {\em Nucleic Acids Res.\@}, 35(Database issue).

\bibitem[Kim and Marcotte, 2008]{Kim2008}
Kim, W.~K. and Marcotte, E.~M. (2008).
\newblock Age-dependent evolution of the yeast protein interaction network
  suggests a limited role of gene duplication and divergence.
\newblock {\em PLoS Comput. Biol.}, 4(11):e1000232.

\bibitem[Kumar et~al., 2006]{KumarKDD2006}
Kumar, R., Novak, J., and Tomkins, A. (2006).
\newblock Structure and evolution of online social networks.
\newblock In {\em Proc.\@ 12th Intl.\@ Conf.\@ on Knowledge Discovery and Data
  mining}, pages 611--617.

\bibitem[Leskovec et~al., 2008]{LeskovecKDD2008}
Leskovec, J., Backstrom, L., Kumar, R., and Tomkins, A. (2008).
\newblock Microscopic evolution of social networks.
\newblock In {\em Proc.\@ 14th Intl.\@ Conf.\@ on Knowledge Discovery and Data
  mining}, pages 462--470.

\bibitem[Leskovec et~al., 2010]{Kronecker}
Leskovec, J., Chakrabarti, D., Kleinberg, J., Faloutsos, C., and Ghahramani, Z.
  (2010).
\newblock Kronecker graphs: An approach to modeling networks.
\newblock {\em J. Mach. Learn. Res.}, 11:985--1042.

\bibitem[Leskovec and Faloutsos, 2007]{Leskovec2007}
Leskovec, J. and Faloutsos, C. (2007).
\newblock Scalable modeling of real graphs using {Kronecker} multiplication.
\newblock In {\em Proc.\@ 24th Intl.\@ Conf.\@ on Machine Learning}, pages
  497--504.

\bibitem[Leskovec et~al., 2005]{Leskovec2005}
Leskovec, J., Kleinberg, J., and Faloutsos, C. (2005).
\newblock Graphs over time: densification laws, shrinking diameters and
  possible explanations.
\newblock In {\em Proc.\@ 11th Intl.\@ Conf.\@ on Knowledge Discovery and Data
  mining}, pages 177--187.

\bibitem[Leskovec et~al., 2007]{LeskovecSDM2007}
Leskovec, J., McGlohon, M., Faloutsos, C., Glance, N., and Hurst, M. (2007).
\newblock Cascading behavior in large blog graphs: Patterns and a model.
\newblock In {\em Proc.\@ 7th SIAM Intl.\@ Conf.\@ on Data Mining}.

\bibitem[Levy and Pereira-Leal, 2008]{Levy2008}
Levy, E.~D. and Pereira-Leal, J.~B. (2008).
\newblock Evolution and dynamics of protein interactions and networks.
\newblock {\em Curr. Opin. Struct. Biol.}, 18(3):349--357.

\bibitem[Li et~al., 2004]{Li2004}
Li, S. et~al. (2004).
\newblock A map of the interactome network of the metazoan \emph{{C.} elegans}.
\newblock {\em Science}, 303(5657):540--543.

\bibitem[Makino et~al., 2006]{Makino2006}
Makino, T., Suzuki, Y., and Gojobori, T. (2006).
\newblock Differential evolutionary rates of duplicated genes in protein
  interaction network.
\newblock {\em Gene}, 385:57--63.

\bibitem[Middendorf et~al., 2005]{Middendorf2005}
Middendorf, M., Ziv, E., and Wiggins, C.~H. (2005).
\newblock Inferring network mechanisms: the \emph{{D}rosophila melanogaster}
  protein interaction network.
\newblock {\em Proc.\@ Natl.\@ Acad.\@ Sci.\@ USA}, 102(9):3192--3197.

\bibitem[Milo et~al., 2002]{Milo2002}
Milo, R., Shen-Orr, S., Itzkovitz, S., Kashtan, N., Chklovskii, D., and Alon,
  U. (2002).
\newblock Network motifs: Simple building blocks of complex networks.
\newblock {\em Science}, 298(5594):824--827.

\bibitem[Mithani et~al., 2009]{Mithani09}
Mithani, A., Preston, G., and Hein, J. (2009).
\newblock A stochastic model for the evolution of metabolic networks with
  neighbor dependence.
\newblock {\em Bioinformatics}, 25(12):1528--1535.

\bibitem[Navlakha et~al., 2009]{NavlakhaJCB2009}
Navlakha, S., Schatz, M.~C., and Kingsford, C. (2009).
\newblock Revealing biological modules via graph summarization.
\newblock {\em J. Comp. Biol.}, 16(2):253--264.

\bibitem[Palla et~al., 2007]{Palla2007}
Palla, G., Barab\a'asi, A.~L., and Vicsek, T. (2007).
\newblock Quantifying social group evolution.
\newblock {\em Nature}, 446(7136):664--667.

\bibitem[Pereira-Leal et~al., 2007]{Pereira-Leal2007}
Pereira-Leal, J.~B., Levy, E.~D., Kamp, C., and Teichmann, S.~A. (2007).
\newblock Evolution of protein complexes by duplication of homomeric
  interactions.
\newblock {\em Genome Biol.\@}, 8(4):R51.

\bibitem[Pereira-Leal et~al., 2006]{Pereira-Leal2006}
Pereira-Leal, J.~B., Levy, E.~D., and Teichmann, S.~A. (2006).
\newblock The origins and evolution of functional modules: lessons from protein
  complexes.
\newblock {\em Philos. Trans. R Soc. Lond. B Biol. Sci.}, 361(1467):507--517.

\bibitem[Shannon et~al., 2003]{Cytoscape}
Shannon, P., Markiel, A., Ozier, O., Baliga, N.~S., Wang, J.~T., Ramage, D.,
  Amin, N., Schwikowski, B., and Ideker, T. (2003).
\newblock Cytoscape: a software environment for integrated models of
  biomolecular interaction networks.
\newblock {\em Genome Res.\@}, 13(11):2498--2504.

\bibitem[Singh et~al., 2007]{Singh2007}
Singh, R., Xu, J., and Berger, B. (2007).
\newblock Pairwise global alignment of protein interaction networks by matching
  neighborhood topology.
\newblock In {\em Proc.\@ 11th Intl.\@ Conf.\@ on Research in Computational
  Molecular Biology (RECOMB)}, pages 16--31.

\bibitem[Tantipathananandh and Berger-Wolf, 2009]{Tantipathananandh2009}
Tantipathananandh, C. and Berger-Wolf, T. (2009).
\newblock Constant-factor approximation algorithms for identifying dynamic
  communities.
\newblock In {\em Proc.\@ 15th Intl.\@ Conf.\@ on Knowledge Discovery and Data
  mining}, pages 827--836.

\bibitem[Tatusov et~al., 2003]{Tatusov2003}
Tatusov, R. et~al. (2003).
\newblock The {COG} database: an updated version includes eukaryotes.
\newblock {\em BMC Bioinformatics}, 4:41.

\bibitem[Vazquez et~al., 2003]{Vazquez2003}
Vazquez, A., Flammini, A., Maritan, A., and Vespignani, A. (2003).
\newblock Modeling of protein interaction networks.
\newblock {\em Complexus}, 1(1):38--44.

\bibitem[Wagner, 2003]{Wagner2003}
Wagner, A. (2003).
\newblock How the global structure of protein interaction networks evolves.
\newblock {\em Proc. Biol. Sci.}, 270(1514):457--466.

\bibitem[Watts and Strogatz, 1998]{Watts1998}
Watts, D.~J. and Strogatz, S.~H. (1998).
\newblock Collective dynamics of `small-world' networks.
\newblock {\em Nature}, 393(6684):440--442.

\bibitem[Wiuf et~al., 2006]{Wiuf06}
Wiuf, C., Brameier, M., Hagberg, O., and Stumpf, M.~P. (2006).
\newblock A likelihood approach to analysis of network data.
\newblock {\em Proc.\@ Natl.\@ Acad.\@ Sci.\@ USA}, 103(20).

\end{thebibliography}
\end{document}